# Emergent 3D Fermiology and Magnetism in an Intercalated Van der Waals System

Luigi Camerano,[1, 2, *] Emanuel A. Martínez,[1] Victor Porée,[3, 4] Laura Martella,[1] Dario Mastrippolito,[5]
Debora Pierucci,[5] Franco D'Orazio,[1] Polina M. Sheverdyaeva,[6] Paolo Moras,[6] Enrico Della Valle,[7]
Tianlun Yu,[7] Moritz Hoesch,[8] Craig M. Polley,[9] Thiagarajan Balasubramanian,[9] Alessandro
Nicolaou,[4] Luca Ottaviano,[1, 2] Vladimir N. Strocov,[7] Gianni Profeta,[1, 2] and Federico Bisti[1]

[1] *Department of Physical and Chemical Sciences,*
*University of L'Aquila, Via Vetoio, 67100 L'Aquila, Italy*
[2] *CNR-SPIN L'Aquila, Via Vetoio, 67100 L'Aquila, Italy*
[3] *Univ Rennes, CNRS, Institut des Sciences Chimiques de Rennes-UMR6226, 35042 Rennes, France*
[4] *Synchrotron SOLEIL, L'Orme des Merisiers, Saint-Aubin, BP 48, F-91192 Gif-sur-Yvette, France*
[5] *Sorbonne Université, CNRS, Institut des NanoSciences de Paris, 4 place Jussieu, 75005, Paris, France*
[6] *CNR-Istituto di Struttura della Materia (CNR-ISM),*
*Strada Statale 14, km 163.5, 34149 Trieste, Italy*
[7] *Swiss Light Source, Paul Scherrer Institute, CH-5232 Villigen PSI, Switzerland.*
[8] *Deutsches Elektronen-Synchrotron DESY, Notkestrasse 85, 22607 Hamburg, Germany*
[9] *MAX IV Laboratory, Lund University, Lund, Sweden*

Intercalation of magnetic atoms into van der Waals materials provides a versatile platform for tailoring unconventional magnetic properties. However, its impact on electronic dimensionality and exchange mechanisms remains poorly understood. Using Fe-intercalated TaS$_2$ as a model system, we combine X-ray absorption and resonant inelastic scattering with angle-resolved photoemission and first-principles calculations to reveal that intercalation reshapes the host electronic structure. We identify a spin-polarized intercalant-host hybridized band with pronounced out-of-plane dispersion crossing the Fermi level, providing an itinerant channel for interlayer magnetic exchange. This mechanism explains the breakdown of a purely atomic picture and establishes a direct link between lattice geometry, electronic dispersion, and magnetic order. Our findings demonstrate that intercalant-induced itinerancy enables tunable interlayer coupling in otherwise layered magnets, offering a general microscopic framework for engineering magnetic dimensionality in a broad class of intercalated vdW materials.

## A. Introduction

The interaction between localized magnetic moments and conduction electrons gives rise to unconventional physical states showing exotic magnetism [1, 2], topological superconductivity [3, 4] and Kondo physics [5, 6]. The intercalation of magnetic atoms in layered metallic van der Waals (vdW) transition metal dichalcogenides (TMDs) (Fig. 1a) provides a multifunctional platform to engineer a variety of quantum states arising from this interaction. Some examples are: highly tunable magnetic states in Fe$_{1/3+x}$NbS$_2$ [7–9], non-coplanar magnetic order giving rise to spontaneous Hall effect in Co$_{1/3}$TaS$_2$ and Co$_{1/3}$NbS$_2$ [10–12], switchable helical spin texture in Ni$_{1/3}$NbS$_2$ [13], chiral helimagnetic states in Cr$_{1/3}$TaS$_2$ and Cr$_{1/3}$NbS$_2$[14–16], tunable Ising ferromagnetism and topological Hall effect in Fe$_x$TaS$_2$ [17, 18], controllable nematic states in Co$_{1/3}$TaS$_2$ [19, 20] and $g$-wave altermagnetism in Co$_{1/4}$NbSe$_2$ [21–24]. All these states are ultimately stabilized through the delicate balance between Kondo and Ruderman-Kittel-Kasuya-Yosida (RKKY) interactions, modulated by the concentration of the transition metal (TM) intercalant, which in turn tunes the doping of the TMD metallic state. A particular interesting platform to study the competition between these interactions is Fe-intercalated TaS$_2$ that exists at different Fe concentrations ($x$) [25–30]. While at very low concentration ($x < 0.05$) Fe intercalation rises the superconducting critical temperature of the host TaS$_2$ [26, 27], for $0.25 < x < 0.40$ it induced ferromagnetism with tunable critical temperatures. Beyond $x > 0.4$ an antiferromagnetic phase emerges [25, 28]. Interestingly, at concentrations $x = 1/4$ and $x = 1/3$, air-stable ordered reconstructions are observed in these compounds [29–32]. In particular, at $x = 1/4$ a $2 \times 2$ reconstruction is observed with a ferromagnetic Curie temperature of $T_c = 135$ K, whereas at $x = 1/3$ the system presents a $\sqrt{3} \times \sqrt{3}$ rotated by 30 degree reconstruction ($\sqrt{3} \times \sqrt{3}R30°$) with a $T_c = 70$ K [25] and large anisotropic magnetoresistance [33, 34]. Additionally, magnetic properties were found to exhibit a pronounced dimensional dependence [35], and signatures of non-ideal 2D behavior have emerged [36] Understanding this rich phase diagram requires a detailed comprehension of how intercalant atoms influence the electronic structure of the TMD, as well as of the role played by the hybridization between the magnetic atoms and the host lattice.

To address these issues, we use Fe$_{1/3}$TaS$_2$ as a model system and combine X-ray absorption spectroscopy (XAS) and resonant inelastic X-ray scattering (RIXS) to probe the local electronic environment, together with angle-resolved photoemission spectroscopy (ARPES) supported by first-principles modeling and multiplet calculations. We report the emergence of an intrinsically three-dimensional, spin-polarized itinerant band induced by intercalation that qualitatively reshapes the electronic structure of the host matrix giving rise to a pronounced $k_z$ dispersion. We discover that this band crosses the Fermi level and provides an efficient channel for mediating interlayer magnetic exchange. Because this mechanism relies only on the presence of metallic van der Waals hosts and symmetry-allowed hybridization with intercalant orbitals oriented along the out-of-plane direction, it is expected to be generic across a broad class of intercalated dichalcogenides. Our results show that intercalant-induced itinerancy acts as a tunable pathway for modifying lattice geometry, electronic dispersion, and magnetic order in van der Waals magnets, offering a unified microscopic perspective for engineering interlayer magnetism.

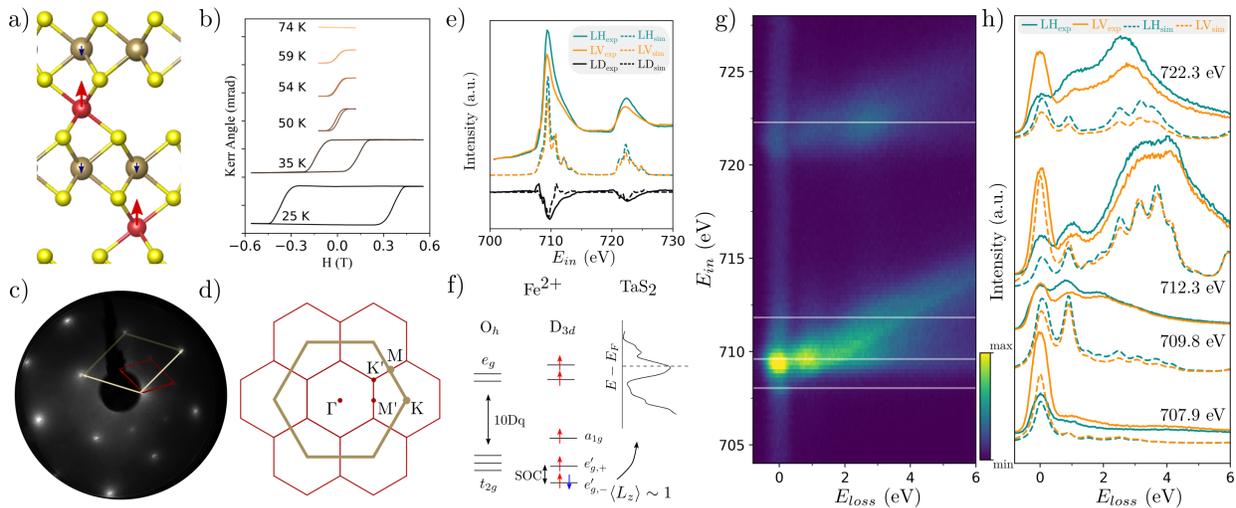

FIG. 1. a) Unit cell of ferromagnetic $Fe_{1/3}TaS_2$ with c-axis aligned along the $z$-direction. The spins on the Fe ions are reported in red, whereas those on the Ta ions are in blue. b) Field dependent MOKE curves at various temperatures of $Fe_{1/3}TaS_2$ using a wavelength of 625 nm. The vertical tick marks are spaced by 10 mrad. c) LEED spots acquired with 200 eV highlighting the $\sqrt{3} \times \sqrt{3}R30°$ reconstruction. d) Sketch of the reciprocal space in which we reported the $\sqrt{3} \times \sqrt{3}R30°$ ordered reconstruction induced by the Fe intercalation. e) Fe $L_{2,3}$ absorption spectra for linearly polarized incident radiation along with the corresponding linear dichroism. The experimental and simulated spectra are displayed in continuous and dashed lines, respectively. f) Energy level scheme showing $O_h$ symmetry breaking into $D_{3d}$ symmetry and the filling of Fe-$d$ orbitals according to the model used to simulate the XAS/RIXS and $TaS_2$ density of states near the Fermi level simulated with HSE06 functional highlighting $Fe^{2+}$ embedded in a metallic matrix. g) RIXS map of the Fe $L_{2,3}$ edges collected using circularly polarized incident radiation. h) Experimental (continuous lines) and simulated (dashed lines) RIXS spectra acquired using linearly polarized incident light. The corresponding incident energies are given above each curve and are indicated by the white horizontal lines in g).

## B. Atomic structure and magnetic order

Bulk $2H$-$TaS_2$ is formed by stacked $1H$-$TaS_2$ monolayers (MLs) in an ABA sequence related by a glide-mirror symmetry (Fig. 1a). The $1H$-$TaS_2$ ML consists of a hexagonal lattice of Ta atoms in trigonal prismatic co-ordination with the chalcogen [37]. Fe intercalation in the interlayer spacing of $2H$-$TaS_2$ induces ferromagnetism (see Fig. 1a). This is demonstrated by our temperature dependent magneto-optical Kerr effect (MOKE) hysteresis loop at various temperature that show a ferromagnetic transition between 59K and 74K, consistent with previous experimental reports [25, 29, 30, 32]. Moreover, the Low Energy Electron Diffraction (LEED) pattern (taken at $E_{kin} = 200$ eV, Fig. 1b) shows presence of $1 \times 1$ spots and an additional periodicity compatible with $\sqrt{3} \times \sqrt{3}R30°$ reconstruction, indicated as a red hexagon, as shown in Fig. 1c where we report a sketch of the reciprocal Brillouin Zone (BZ) of both the unit cells. We indicate with prime letters the BZ of the superstructure. This superstructure is typical of the $x = 1/3$ concentration of Fe in pristine $2H$-$TaS_2$.

## C. XAS and RIXS: limits of a purely atomic picture

XAS and RIXS provide direct access to the local electronic configuration of the Fe intercalant and represent a natural starting point for understanding the magnetic properties of $Fe_{1/3}TaS_2$. In Fig. 1e-f-g-h we present polarization-dependent XAS spectra at the Fe $L_{2,3}$ edges together with momentum-integrated RIXS measurements, supported by many-body multiplet cluster calculations and density functional theory calculations (DFT) (see Supplementary Information (SI) for further details).

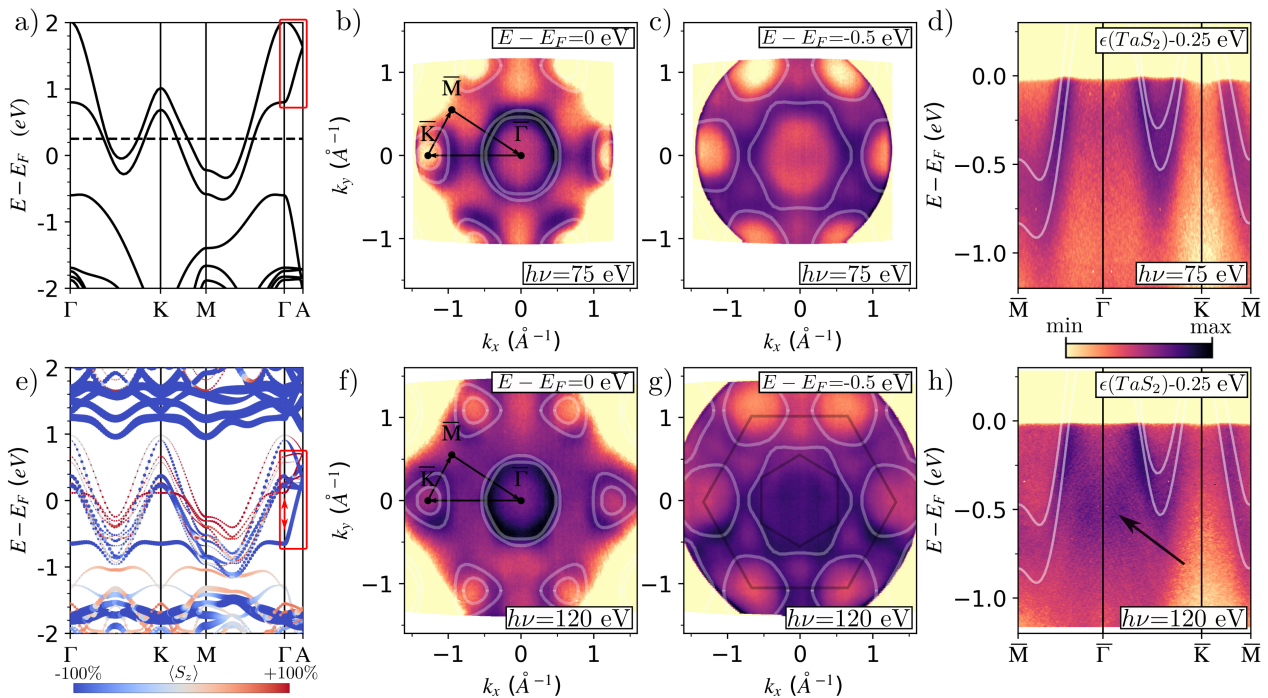

FIG. 2. a) and e) show the band structure of $TaS_2$ and $Fe_{1/3}TaS_2$ respectively. In a) the dashed line at 0.25 eV indicates the energy shift applied to account for Fe doping (see main text). In e) the size of the points represents the Fe-$d$ character of the bands while the colormap is $\langle S_z \rangle$.b) and f) Fermi surface map acquired at $h\nu = 75$ eV and $h\nu = 120$ eV respectively, indicating the high-symmetry directions used to define the band structure path. c) - g) Constant energy cuts at $E - E_F = -0.5$ eV at $h\nu = 75$ eV and $h\nu = 120$ eV respectively. c) - f) Interpolated band structure along the path defined in a) -d). We superimposed the band structure of $TaS_2$ calculated with HSE06 functional shifted by 0.25 eV accounting for the Fe electron doping. The black arrows in d) - h) indicate an additional band not present in the $TaS_2$ band structure.

The shape of the XAS and RIXS spectra suggests a $2^+$ oxidation state of the Fe ion, a claim that is further corroborated by our models. By acquiring the XAS with different polarizations (horizontal LH, vertical LV, see Fig. 1e and Fig. S2 for the details of the geometry), we observe a linear dichroism (LD) in the Fe $L_{2,3}$ XAS spectra (Fig. 1e), signaling a deviation from cubic symmetry. This behavior is consistent with a trigonal distortion of the Fe octahedral environment, as expected from the crystallographic structure of the intercalated compound. The overall line shape and energy position of the absorption edges are well reproduced by multiplet simulations assuming a $Fe^{2+}$ configuration in a $D_{3d}$ point group crystal field, indicating a nominal $3d^6$ electronic filling. The corresponding orbital level scheme is shown in Fig. 1f, where the splitting of the $3d$ manifold reflects the combined effects of crystal field and spin–orbit coupling. RIXS measurements across the Fe $L_{2,3}$ edges further constrain the local electronic structure. The RIXS intensity map (Fig. 1g) is dominated by two main loss features at approximately 1 eV and 3 eV, whose resonance be-havior and polarization dependence are characteristic of crystal-field–allowed $d$–$d$ excitations. As shown in Fig. 1h, these excitations are quantitatively reproduced by the same multiplet model used for the XAS, confirming the internal consistency of the atomic description (the details of the many-body multiplet cluster model used to model the spectra, sketched in Fig. 1f, is provided in SI). The lower-energy excitation corresponds predominantly to transitions into the $e'_g$ manifold, while the higher-energy loss arises from a dense set of excited states overlapping with the fluorescence background (see also Fig. S1 panel c for a visualization of the energy levels). The polarization dependence observed both in XAS and RIXS indicates a ground state characterized by a sizable orbital moment, stabilized by spin–orbit coupling within the $e'_g$ doublet. This is consistent with previous reports of strong magnetic anisotropy in Fe-intercalated $TaS_2$ compounds [18, 33]. Using the same multiplet parameters, we calculate the local magnetic response and find a large out-of-plane anisotropy [18, 34, 37], in agreement with magneto-optical Kerr effect measurements (Fig.1b). However, despite the success of the atomic multiplet description in reproducing the main spectral features, important discrepancies emerge when comparing the local picture with experiments. In particular, the calcu-lated saturation magnetization $m_{sat} \simeq 5$ $\mu_B$ (see Fig. S1 panels a and b), that is the result of an electronic configuration with $\langle S_z \rangle \simeq 4$ $\mu_B$ and $\langle L_z \rangle \simeq 1$ $\mu_B$, exceeds the experimentally measured value $m_{sat} \simeq 4$ $\mu_B$ [33, 36, 38, 39], and the RIXS spectra do not display a clear excitation gap that would be expected for a purely Fe localized electronic configuration. Instead, the spectral weight is broadened and partially overlaps with the elastic peak, a behavior more typical of itinerant or mixed localized–itinerant systems. These deviations signal the breakdown of a purely atomic description and point to a finite hybridization between Fe-3$d$ states and the TaS$_2$ host lattice. While XAS and RIXS remain dominated by local excitations, they carry indirect fingerprints of this hybridization through the reduced effective magnetic moment and the absence of a well-defined insulating gap between Fe-$d$ states. This behavior is representative of an itinerant system with the persistence of localized state typical of intercalated TMDs but it could be also visible in intrinsic metallic magnets as Fe$_3$GeTe$_2$ [40] and doped insulating magnets as CrGeTe$_3$ [41, 42].

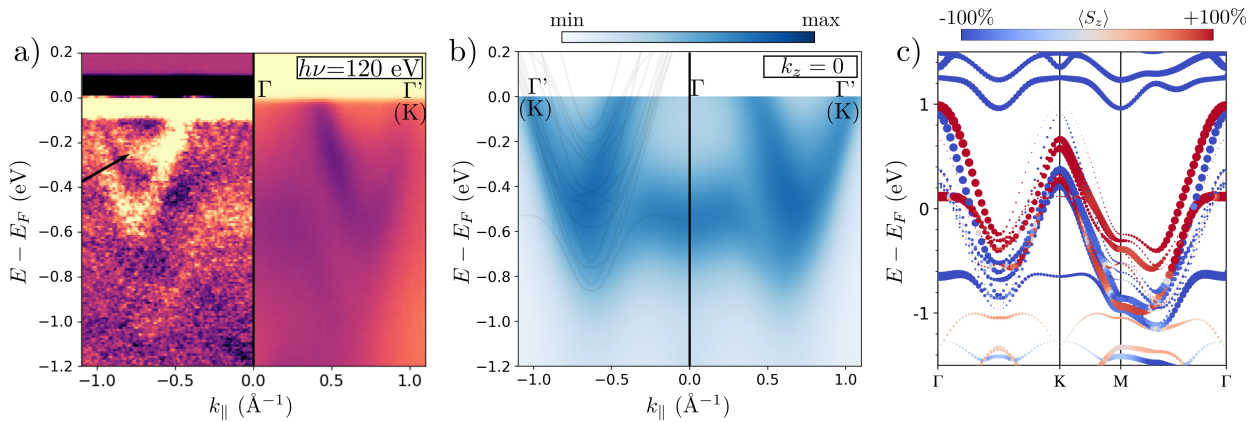

FIG. 3. a) Energy dispersion along the $\Gamma - K$ line acquired at $h\nu = 120$ eV. The left panel report the second derivative of the right panel plot. b) Simulation of the ARPES intensity (See Methods for further details). On the left, we superimposed the band structure calculated using the PBE+$U$+$SOC$ functional, shifted by 0.1 eV to account for an overall p-doping in the sample. c) Unfolded and spin-polarized band structure. The size of the point are proportional to the weight of the unfolding, while the colormap is for the spin polarization $\langle S_z \rangle$.

## D. DFT and ARPES: evidence for hybridization

While multiplet calculations correctly capture the local crystal-field physics of the Fe ions, they do not account for the itinerant electronic degrees of freedom introduced by intercalation. Density functional theory naturally incorporates this itinerancy and allows us to explicitly resolve the hybridization between Fe-$d$ and Ta-$d$ states, resulting in a metallic ground state, which is absent in multiplet calculations. DFT calculations on Fe$_{1/3}$TaS$_2$ (see SI for further details) correctly capture this hybridization, resulting in a $\langle S_z \rangle = 3.5$ $\mu_B$ and $\langle L_z \rangle = 0.7$ $\mu_b$. The reduced saturation magnetization obtained is due to the partial delocalization of the Fe-$d$ electrons induced by hybridization with Ta-$d$ states, which transfers spin density away from the Fe site and naturally lowers the local moment.

Inspecting the band structure (see Fig. 2a-e for a comparison of the band structure. In Fig. 2e the Fe$_{1/3}$TaS$_2$ band structure is projected on Fe-$d$ orbitals and the colormap correspond to the $S_z$ expected value) we note that low dispersive spin-minority states are located at $E - E_F = $-1.8 eV and above $E - E_F = 1$ eV in conduction band. The dispersion along the out-of-plane momentum direction is strongly modified by the intercalated ion. In particular, we found a strong Fe-$d$-Ta-$d$ hybridized spin-minority states that crosses the Fermi level along the $\Gamma A$ direction, as highlighted by the red boxes in the Fig 2e. The pronounced dispersion of this hybridized band along $k_z$ implies a finite out-of-plane hopping amplitude, demonstrating that the corresponding electronic states are intrinsically three-dimensional rather than confined to individual TaS$_2$ layers.

Indeed, we found that the Fe ions induce a small antiferromagnetically coupled moment on the nearest Ta atoms ($\sim$ -0.04 $\mu_B$). Moreover, analyzing the occupation matrix in our first principle calculations (see SI for further details) we found that the glide mirror symmetry of the host compound imposes a different local orientation of the Fe in-plane orbitals. This peculiar property of intercalated TMD is ultimately the origin of the emergence of alternmagnetic phases in this class of compounds when A-type antiferromagnetism is present [21–24]. Therefore, our first-principle calculations show that beyond the presence of localized Fe-$d$ states, responsible for the magnetic properties of the compound, the out-of-plane energy dispersion is significantly modified by the ion intercalation, resulting in an additional dispersive band crossing the Fermi level.

Evidence for Fe-$d$–Ta-$d$ hybridization was obtained from ARPES measurements, shown in Fig.2. In particular, we report ARPES spectra acquired with photon energy h$\nu$ = 75 eV in panels b-c-d and h$\nu$ = 120 eV in panels f-g-h at T = 15 K. We superimpose DFT calculation of the pristine 2$H$-TaS$_2$ along the M$\Gamma$KM path shifted by 0.25 eV to account for electron doping induced by the TM intercalant atom and better compare with the electronic structure of the host compound (the $k_z$ dispersion mainly affect the broadening in the M point [43]). At h$\nu$ = 75 eV, see Fig. 2b-c, the isoenergy maps along the plane $\bar{\Gamma}\bar{K}\bar{M}$ highlight the presence of a hole pocket at $\bar{K}$ similar to the one present in 2$H$-TaS$_2$ [43–45].

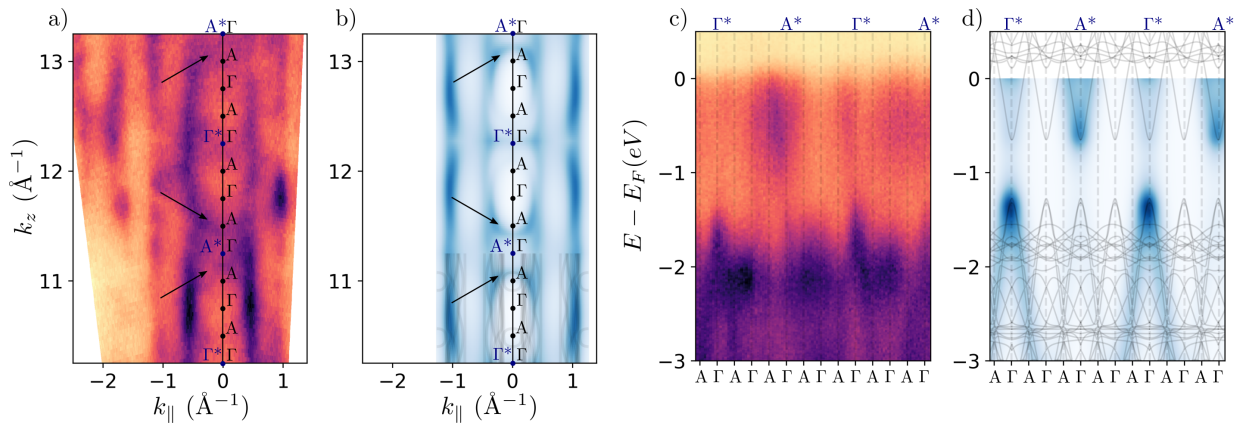

FIG. 4. a) Soft X-rays ARPES Fermi surface map in the $\Gamma AK$ plane using CR polarized light. We reported in black several BZ along the $k_z$ momentum. In dark blue we reported symmetry points stemming for the observed periodicity in the ARPES spectrum. The black arrows indicate the Fermi surface lines not present in the TaS$_2$ spectrum. b) ARPES intensity simulation of the $\Gamma AK$ plane. For the intensity simulations we used Fe-$d$ and Ta-$d$ complex projections. In the bottom, we reported the Fermi surface as seen from the bare band structure. c) Band dispersion along the $\Gamma A$ direction. d) ARPES intensity simulation of the $\Gamma A$ direction. In black, we superimpose $k_z$ dispersion of the band at $(k_x, k_y) = (0, 0)$.

In Fig. 2d we report the band structure along the $\bar{M}\bar{\Gamma}\bar{K}\bar{M}$ lines of the surface BZ, showing Ta$-d$ bands crossing the Fermi level and no significant difference with respect to pristine 2$H$-TaS$_2$ apart from a substantial rigid doping and diffuse spectral weight at $\bar{\Gamma}$. Changing the photon energy, h$\nu$ = 120 eV, novel features are visible. Indeed, the energy cut at $E - E_F$ = -0.5 eV show signal near the BZ center, where no bands of pristine 2$H$-TaS$_2$ are present [43]. These bands are visible in Fig.2h where the band structure along the $\bar{M}\bar{\Gamma}\bar{K}\bar{M}$ is reported. Our first principle calculations on Fe$_{1/3}$TaS$_2$ show that these bands are hybridized Fe-$d$ and Ta-$d$ bands (see Fig. 2e) dispersing in the $\bar{\Gamma}\bar{K}\bar{M}$ plane.

For improving the comparison of DFT calculations with experimental measurement we report in Fig. 3 high-statistic measurement along $\Gamma$K parallel direction compared with intensity simulation taking into account interference effects from initial wavefunction (see SI for further detail). We further show the second derivative of the intensity in the left panel of Fig.3a to better resolve the splitting of the bands. In the left panel of Fig. 3b we report band structure of the compound in the Fe$_{1/3}$TaS$_2$ unit cell.

We note that Ta-$d$ bands that in 2$H$-TaS$_2$ were completely empty by charge density wave distortion [43, 44,

46] are now filled and reach $E - E_F \sim$ -0.3 eV (as highlighted by black arrow in the Fig. 3a). The observed band broadening with respect to pristine 2$H$-TaS$_2$ is attributed to Fe intercalation, which introduces hybridization between Fe-$d$ and Ta-$d$ states and local distortion redistributing the spectral weight of the folded Ta-$d$ bands. This feature is correctly captured by our intensity simulation as shown in Fig. 3b-c. Furthermore, our simulations fully describe the vanishing spectral weight observed in the experiment of the Fe-$d$ derived band in $\bar{\Gamma}'$ of the $\sqrt{3} \times \sqrt{3}R30°$ BZ as shown in Fig. 3b. In these simulations, the initial state in the matrix element $M$ is linear combination of Fe-$d_{z^2}$, Fe-$d_{x^2-y^2}$, Fe-$d_{xz}$ and Ta-$d_{z^2}$, Ta-$d_{x^2-y^2}$, Ta-$d_{xz}$ accounting for the LH polarization of the light. The breaking of time reversal symmetry and centrosymmetry by intercalation of ferromagnetic aligned Fe ions naturally leads to analyze the spin splitting of the measured bands. In Fig. 3c and in Fig. S6, we show the unfolded band structure (from the $\sqrt{3} \times \sqrt{3}R30°$ unit cell to the 2H-TaS$_2$ unit cell) at different $k_z$, where the colormap represents the spin polarization along the $z$ axis ($\langle S_z \rangle$). The spin degeneracy that, in the pristine system, arises from the compensation of Ising spin splitting due to glide mirror symmetry between layers is now lifted. The hybridization between the Fe local magnetic moments and the Ta conduction states induces a pronounced spin splitting of the Ta-$d$ bands, effectively acting as a Zeeman field along $z$ on otherwise spin-degenerate states. This hybridization arises from a Kondo-like Hamiltonian

$$H_{dd} = J_{dd} \sum_i \mathbf{S}_i \cdot \mathbf{s}(\mathbf{r}_i) \qquad (1)$$

where the itinerant electrons (Ta$_{d_{z^2}}$ like) with spin-density $\mathbf{s}(\mathbf{r}_i)$ interact with localized Fe-$d$ spins $S_i$. Thus, the induced spin splitting of the Ta-derived bands reflects an effective exchange field, $\Delta_{ex} \sim J_{dd}$, generated by the Fe moments $S_i$, indicating that the itinerant electrons actively participate in the magnetic coupling, with a strength $J_{dd}$.

The splitting is significantly larger in the ΓKM plane than in the AHL plane; for instance, at the Γ point the spin-minority band is occupied while the spin-majority band remains unoccupied, enabling the observation of this spin-polarized band at the Γ point in the ARPES spectra, as clearly visible in Fig. 3a-b.

Taken together, DFT and ARPES demonstrate that Fe intercalation generates a spin-polarized, $k_z$-dispersive itinerant band that qualitatively alters the electronic dimensionality of the system, manifesting as a previously overlooked spin-polarized band [18, 47], and provides the microscopic basis for interlayer magnetic coupling.

The presence of such hybridized states, can, in principle, mediate the coupling between different 2D magnetic Fe lattices, changing the magnetic behavior from 2D-like to 3D. Thus, to probe the 3D dispersion along the out-of-plane momentum direction we carried out soft-X-rays ARPES (SX-ARPES) with photon energies from 350 eV to 700 eV and circular right (CR) polarization, reporting the corresponding spectra in Fig. 4 (see SI for further detail where we report ARPES spectra acquired with circular right (CL) polarization). In Fig. 4a-b we show SX-ARPES Fermi surface map in the Γ$AK$ plane and the corresponding first-principles DFT simulation of the spectrum.

The first important evidence is the presence of a three dimensional Fermi surface topology, showing closed 3D sheets, as evidenced by the pronounced features indicated by the black arrows in Fig. 4a-b. On the contrary, these features are completely absent in the host compound [43] (see also Fig. S7 for a comparison of the measured out-of-plane band dispersion in Fe$_{1/3}$TaS$_2$ and calculations in TaS$_2$), which shows a 2D cylindrical Fermi surface. Our first-principle calculation, see Fig.4-b, confirms that these states originate from the out-of-plane hybridization between Ta-$d_{z^2}$ and Fe-$d_{z^2}$, which cross the Fermi level along the Γ − A direction with a linear dispersion (see Fig. 4d), while the dispersion of this band along the in-plane momentum shows a nearly zero velocity at Γ suggesting a pronounced electronic anisotropic transport behavior (see Fig. S4 and S5 where we report dispersion in the ΓK with different photon energy and $k_z$s going from Γ to A). To properly describe such band, we first note that the measured Fermi surface map shows a $k_z$ periodicity four times larger than the one of the lattice of Fe$_{1/3}$TaS$_2$ (we introduced the Γ* and A* points to describe the observed periodicity). This multiple doublings of the signal cannot be captured by band structure calculations, reported in the bottom of Fig.4b, which by definition has the periodicity of the BZ unit cell. We ascribe this experimental evidence to a consequence of quantum interference effects between atoms belonging to different layers combined with the geometric effect of intercalant Fe ion. These interference effects have been already observed in pristine 2H-TaS$_2$ [43] and

2H-NbS$_2$ [48] and arise from the phase mismatching in different sublattice sites of the initial state wavefunction (different TMD layers). In the latter cases, the period is doubled. In this case, the already doubled period is doubled again due to the intercalation of Fe ions in the vdW gap of 2H-TaS$_2$ that induces an additional contribution to the initial Bloch wavefunction's phase, whose impact on the intensity of the SX-ARPES signal can be measured. To better clarify the origin of this pure quantum effect induced by the Fe intercalants, we construct one dimensional (1D) models to capture the out-of-plane dispersion in the pristine and intercalated phases (see SI and Fig. S3). While the $k_z$ dispersion in pristine 2H-TaS$_2$ is well captured by a monoatomic chain model, Fe$_{1/3}$TaS$_2$ requires two inequivalent atomic sites. Non-magnetic band-structure calculations (see Fig. S3) and relative 1D model elucidate the underlying mechanism: at the Γ$_1$ point, the wavefunctions on sublattices A and B acquire a relative π-phase shift, in contrast to Γ$_0$ and Γ$_2$, where they remain in phase. The same behavior is observed in both measured and simulated ARPES spectra, where the spectral weight at Γ* differs from that at A* due to phase mismatch in the initial-state wavefunctions induced by the intercalant atoms. Therefore, by comparing SX-ARPES and first principle-calculation we unveil the out-of-plane dispersion of previously overlooked band induced by the Fe intercalant atom in the vdW gap of TaS$_2$. Moreover, apart from the methodology we used to reconstruct the signal, it is important to note that such strong interference effects is reflecting once more the relevance of the Fe intercalation in shaping the band structure of the material (see also Fig. S7 for a comparison with the $k_z$ dispersion of the pristine 2H-TaS$_2$).

### E. The role of interlayer coupling

To understand the role of this band in shaping the magnetic properties of the material, we calculated the interlayer magnetic coupling by total energy difference method as a function of the out-of-plane lattice parameter $c$. Interestingly, the calculated value for the interlayer coupling is one order of magnitude larger than the one calculated for paradigmatic bulk 2D magnet CrI$_3$ [49]. We also stabilize a higher-energy phase with $\langle L_z \rangle \simeq 0 \mu_B$ by explicitly controlling orbital occupancy [50–52] and mixing parameter in the self-consistent cycle; in this case, the interlayer coupling changes sign, yielding an antiferromagnetic ground state. This underscores the necessity of stabilizing the correct orbital configuration to properly capture the magnetic ground state, which is otherwise missed when the orbital degree of freedom is not accounted for [53].

The magnetic coupling between Fe atoms in each layer are effectively coupled through the dispersive out-of-plane band, in an anisotropic RKKY-like mechanism as described by Eq. (1). As a result, the effective interlayer magnetic coupling $J_z$ depend on the spin-susceptibility $\chi^\alpha(\mathbf{r})$ and scales with the square of the intercalant–electron exchange $J_{dd}$ as $J_z^\alpha \sim J_{dd}^2 \chi^\alpha$ (where $\alpha$ is for the spin anisotropy). Writing $\chi^\alpha(\mathbf{r})$ in momentum space and focusing on the out-of-plane component we obtain:

$$\chi^\alpha(q_z) = \sum_k \frac{f(\epsilon_k) - f(\epsilon_{k+q_z})}{\epsilon_{k+q_z} - \epsilon_k} |M_{S^\alpha}|^2 \qquad (2)$$

where $f(\epsilon_k)$ is the Fermi-Dirac distribution and $M_{S^\alpha}$ is given by $\langle k + q_z | S^\alpha | k \rangle$. Thus, the absence of out-of-plane electronic dispersion, states connected by a finite $q_z$ are nearly degenerate, prevents real electronic transitions and suppresses the spin susceptibility along the interlayer direction. The emergence of a $k_z$-dispersive itinerant band therefore activates a finite interlayer spin response, whose magnitude increases with the band's curvature along the out-of-plane direction, namely the hopping amplitude, $t^*$. Indeed, we estimate the effective hopping $t^*$ by expanding near the $\Gamma$ point the first-principle energy dispersion $\epsilon(k_z) = a_1 + t^* c^2 k_z^2$. In Fig.5 we show the evolution of these quantities as a function of the out-of-plane lattice parameter $c$. Reducing $c$ systematically enhances the interlayer exchange $J_z$, revealing the sensitivity of the magnetic coupling to the interlayer spacing. In parallel, the effective hopping $t^*$ of the out-of-plane dispersive band exhibits an increase. Moreover, the antiferromagnetically aligned moment on the otherwise non-magnetic Ta site grows in absolute value with increasing $J_z$, further indicating that the interlayer exchange is mediated by hybridization Fe-$d$-Ta-$d$ bands.

Therefore the emergence of a $k_z$ dispersive Fe–Ta hybridized band provides the microscopic origin of the interlayer exchange, with $J_z$ increasing with the out-of-plane hopping amplitude. Reducing the interlayer spacing enhances the out-of-plane hopping of the itinerant band, which in turn strengthens the interlayer magnetic exchange. This concurrence demonstrates that the hybridized itinerant electrons contributing to the $k_z$-dispersive band provide a channel for mediating $J_z$, establishing a direct and tunable link between lattice geometry, electronic dispersion, and interlayer magnetic exchange.

## F. Conclusion

In conclusion, we have shown that magnetic intercalation in metallic van der Waals materials qualitatively reshapes the electronic structure by generating itinerant, $k_z$ dispersive states that fundamentally alter the nature of magnetic coupling. Our analysis demonstrates that a purely atomic picture cannot account for the magnetic behavior of this compound. Instead, we reveal the essential role of hybridization between the magnetic Fe intercalants and the metallic $TaS_2$ host, manifested in a strongly three-dimensional dispersive band and a reduced magnetization at saturation.

The emergence of a spin-polarized Fe–Ta hybridized band crossing the Fermi level provides an efficient itinerant channel for interlayer spin propagation. This establishes a direct microscopic link between the electronic dispersion and three-dimensional magnetic order, enabling a tunable out-of-plane exchange interaction that ultimately gives rise to A-type antiferromagnetism in intercalated altermagnets [21–24], as well as to a tunable helical spin texture [13]. These findings uncover the microscopic mechanism behind the divergence from purely two-dimensional magnetism in this system and provide a foundation for understanding and engineer the emergence of exotic magnetic phases in transition metal-intercalated dichalcogenides.

## I. DATA AVAILABILITY STATEMENT

All data that support the findings of this study are included within the article and supplementary materials.

## II. ACKNOWLEDGEMENTS

This work was funded by the European Union-NextGenerationEU under the Italian Ministry of University and Research (MUR) National Innovation Ecosystem Grant No. ECS00000041 VITALITY-CUP E13C22001060006. F.B. acknowledges funding from the National Recovery and Resilience Plan (NRRP), Mission 4, Component 2, Investment 1.1, funded by the European Union (NextGenerationEU), for the project "TOTEM" (CUP E53D23001710006 - Call for tender No. 104 published on 2.2.2022 and Grant Assignment Decree No. 957 adopted on 30/06/2023 by the Italian Ministry of Ministry of University and Research (MUR)) and for the project "SHEEP" (CUP E53D23018380001 - Call for tender No. 1409 published on 14.9.2022 and Grant Assignment Decree No. 1381 adopted on 01/09/2023 by the Italian Ministry of University and Research (MUR)). We acknowledge Elettra Sincrotrone Trieste for financial support under the SUI internal project and for providing access to its synchrotron radiation facilities (proposal number 20240507). We acknowledge the MAX IV Laboratory for beamtime on the Bloch beamline under proposal 20241631. Research conducted at MAX IV, a Swedish national user facility, is supported by Vetenskapsrådet (Swedish Research Council, VR) under contract 2018-07152, Vinnova (Swedish Governmental Agency for Innovation Systems) under contract 2018-04969 and Formas under contract 2019-02496. D.P. and D.M were supported by a public grant overseen by the French National Research Agency (ANR) through the grants Operatwist (No. ANR-22-CE09-0037-01), E-map (No. ANR-23-CE50-0025), and as part of the "Investissements d'Avenir" program (Labex NanoSaclay, reference: ANR-10-LABX-0035) and from the CNRS through the MITI interdisciplinary programs (project WITHIN). V. P. and A. N. were supported by a public grant overseen by the French National Research Agency (ANR) as part of the TRIXS project ANR-19-CE30-0011, and by the BONASPES project (ANR- 19-CE30-0007) as well as the IMPRESS project that has received funding from the HORIZON EUROPE framework program for research and innovation under Grant Agreement No. 101094299.

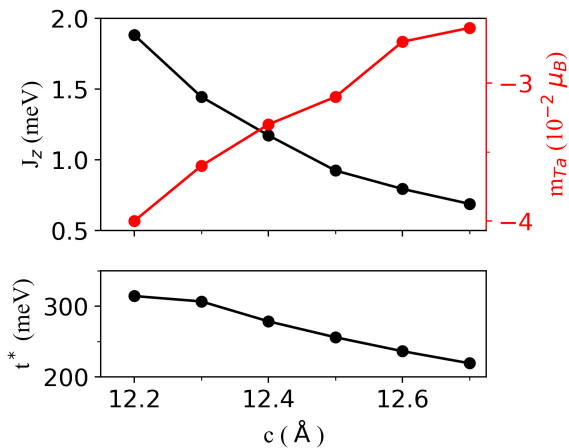

FIG. 5. Magnetic interlayer coupling $J_z$ and estimated effective hopping $t^*$ of the out-of-plane dispersive band crossing the Fermi level as a function of the lattice parameter $c$.

Supplemental material and supporting information for

# Emergent 3D Fermiology and Magnetism in an Intercalated Van der Waals System

Luigi Camerano[a,b], Emanuel A. Martínez [b], Victor Porée[c,d], Laura Martella [a], Dario Mastrippolito [e], Debora Pierucci [e], Franco D'Orazio [a], Polina M. Sheverdyaeva [f], Paolo Moras [f], Enrico Della Valle [g], Tianlun Yu [g], Moritz Hoesch [h], Craig M. Polley [i], Thiagarajan Balasubramanian [i], Alessandro Nicolaou [d], Luca Ottaviano [a,b], Vladimir N. Strocov [g], Gianni Profeta [a,b], Federico Bisti [a]

[a] Department of Physical and Chemical Sciences, University of L'Aquila, Via Vetoio 67100 L'Aquila, Italy
[b] CNR-SPIN L'Aquila, Via Vetoio, 67100 L'Aquila, Italy c/o Department of Physical and Chemical Sciences, University of L'Aquila, Via Vetoio, 67100 L'Aquila, Italy
[c] Univ Rennes, CNRS, Institut des Sciences Chimiques de Rennes-UMR6226, 35042 Rennes, France
[d] Synchrotron SOLEIL, L'Orme des Merisiers, Saint-Aubin, BP 48, F-91192 Gif-sur-Yvette, France
[e] Sorbonne Université, CNRS, Institut des NanoSciences de Paris, 4 place Jussieu, 75005, Paris, France
[f] CNR-Istituto di Struttura della Materia (CNR-ISM), Strada Statale 14, km 163.5, 34149 Trieste, Italy
[g] Swiss Light Source, Paul Scherrer Institute, CH-5232 Villigen PSI, Switzerland.
[h] Deutsches Elektronen-Synchrotron DESY, Notkestrasse 85, 22607 Hamburg, Germany
[i] MAX IV Laboratory, Lund University, Lund, Sweden

## CONTENTS

# I. THEORETICAL SIMULATIONS

Density functional theory calculations were performed using the Vienna ab-initio Simulation Package (VASP) [1, 2], using the generalized gradient approximation (GGA) in the Perdew-Burke-Ernzerhof (PBE) parametrization for the exchange-correlation functional [3], including SOC. Interactions between electrons and nuclei were described using the projector-augmented wave method. Energy thresholds for the self-consistent calculation was set to $10^{-6}$ eV and force threshold for geometry optimization $10^{-4}$ eV $\text{Å}^{-1}$. For the calculation of interlayer magnetic exchange the energy thresholds for the self-consistent calculation was set to $10^{-7}$. A plane-wave kinetic energy cutoff of 550 eV was employed. The Brillouin zone was sampled using an $8 \times 8 \times 4$ Gamma-centered Monkhorst-Pack grid. To account for the on-site electron-electron correlation on localized Fe-$d$ orbitals we used the GGA+U approach with an effective Hubbard term $U = 4.42$ eV as we calculated with linear response theory [4]. The 2H-TaS$_2$ lattice parameter are set to the experimental ones: $a = 3.31$ Å and $c = 12.07$ Å [5], while for the Fe$_{1/3}$TaS$_2$ $a = 5.737$ Å and $c = 12.28$ Å [6, 7].

For the band structure of 2H-TaS$_2$ in Fig. 3 a-d-h we used HSE06 functional [8–10] to better describe bands around the M point [11], using the FHI-AIMS simulation package [12, 13], which is an accurate all-electron full-potential electronic structure package based on numeric atom-centered orbitals, with so-called "tight" computational settings. The screened hybrid functional HSE06 with the mixing factor $\alpha = 0.25$ and screening parameter $\omega = 0.11$ Bohr$^{-1}$ was used for the exchange-correlation energy.

Due to the localization of Fe-$d$ orbitals, different metastable phases can be stabilized for Fe$_{1/3}$TaS$_2$. To stabilize large orbital moment phase we used a mixing parameter $\alpha_{mix} = 0.22$ and the occupation matrix control as implemented in VASP [14]. We notice that this phase is indeed the ground state of the system by comparing the total energy.

ARPES intensity simulations are based on the standard model [15–17], where the Lorentzian spectral function is weighted with the square modulus of the matrix element:

$$I(\mathbf{k}, E) = \frac{\sigma}{(E - E_{\mathbf{k}})^2 + \sigma^2} |M|^2 \,, \tag{1}$$

where $E_{\mathbf{k}}$ are the first principle computed eigenvalues, $\sigma$ is the spectral width. $|M|$ is an adapted version of the general $M = \psi_f |\mathbf{A} \cdot \mathbf{p}| \psi_i$ in a DFT framework, that takes into account interference from initial state weighted as $M \sim \sum_j \pm C_j^{n\mathbf{k}}$, where $j$ stand for sublattice indices, $C_j^{n\mathbf{k}} = Y_j |\psi_{\mathbf{k}n}^{KS}$ is the projection of the Khon-Sham wavefunction on suitable atomic orbitals $Y_j$ and $\pm$ fix the symmetry of the initial state according to the polarization of the light [11, 18, 19]. The coefficients $\tilde{C}_j^{n\mathbf{k}}$ contains the information about interference effects.

The relative rotation of the Fe-$d$ orbitals due to the glide mirror symmetry is analyzed using the $d$ density matrix for the down spinon component $n_{Fe-d}^{\downarrow}$. Defining $U$ as the rotation matrix for the cubic harmonics we can write $n_{Fe_1-d}^{\downarrow} = U(\phi = 60°) \, n_{Fe_2-d}^{\downarrow} U^{\dagger}(\phi = 60°)$.

The quantum many body script language Quanty [20–22] was used to reproduce the XAS and RIXS spectra considering a crystal electric field model. The local reference frame used to express the incident and outgoing polarization is represented in Fig. S2b. Slaters integrals, $F_{2,4}$ and $G_{1,3}$ were rescaled to 75 % and 65 % of their Hartree-Fock values, respectively. The spin-orbit coupling was also scaled down by 50 %. In order to account for the linear dichroism observed in the absorption spectra, the $D_{3d}$ point group was considered to capture the crystal-electric field experienced by the Fe$^{2+}$ ions. In this point group, the five $3d$ orbitals split into 3 levels : one non-degenerate ($a_{1g}$) and two doubly degenerated ($e_g$ and $e'_g$) orbitals. Based on the DFT calculations and observed dd transitions, the following orbital energies were used: $e'_g$ = -0.2 eV, $a_{1g}$ = 0 eV, $e_g$ = 0.7 eV.

# II. EXPERIMENTAL METHODS

## A. MOKE methods

MOKE measurements were performed with an homemade experimental setup using laser with 625 nm wavelength. MOKE rotation angles was measured as a function of the out-of-plane magnetic field (aligned with the crystallographic $c$ axis), in the range ±5600 Oe. The bulk samples were cleaved before the MOKE measurements, which were carried out in a cryostat at a base pressure of $1 \times 10^{-6}$ mbar.

## B. XAS-RIXS methods

X-ray absorption spectroscopy and resonant inelastic x-ray scattering measurements at the Fe $L_{2,3}$ edges were performed on the inelastic branch of the SEXTANTS beamline of the synchrotron SOLEIL [23]. RIXS spectra were acquired employing the AERHA spectrometer [24] with a fixed scattering angle of $2\theta = 85°$ and an overall resolution of 570 meV, while XAS spectra were obtained via total electron yield with a resolution of 120 meV. All measurements were carried out at room temperature on a freshly cleaved millimeter-size single crystal using the MAGELEC sample environment [25]. The latter was glued to a copper Omicron-type sample holder using silver paint, with its $a$ crystallographic axis along the vertical experimental direction (i.e. parallel to the vertical beam polarisation LV) and its $c$ crystallographic axis within the scattering plane (i.e. the horizontal experimental plane). XAS and RIXS spectra were acquired using linear (vertical LH, horizontal LV) polarisation in both grazing and normal geometry (with incident angle of 20° and 80°, respectively). The 20° RIXS data are depicted in Fig. S1d, while the 80° data are shown in Fig. S1e. A scheme of the experimental geometry is given in Fig. S2a. Additionally, a RIXS map covering the Fe $L_{2,3}$ energy range was collected in grazing incidence, using a circularly polarised incident radiation.

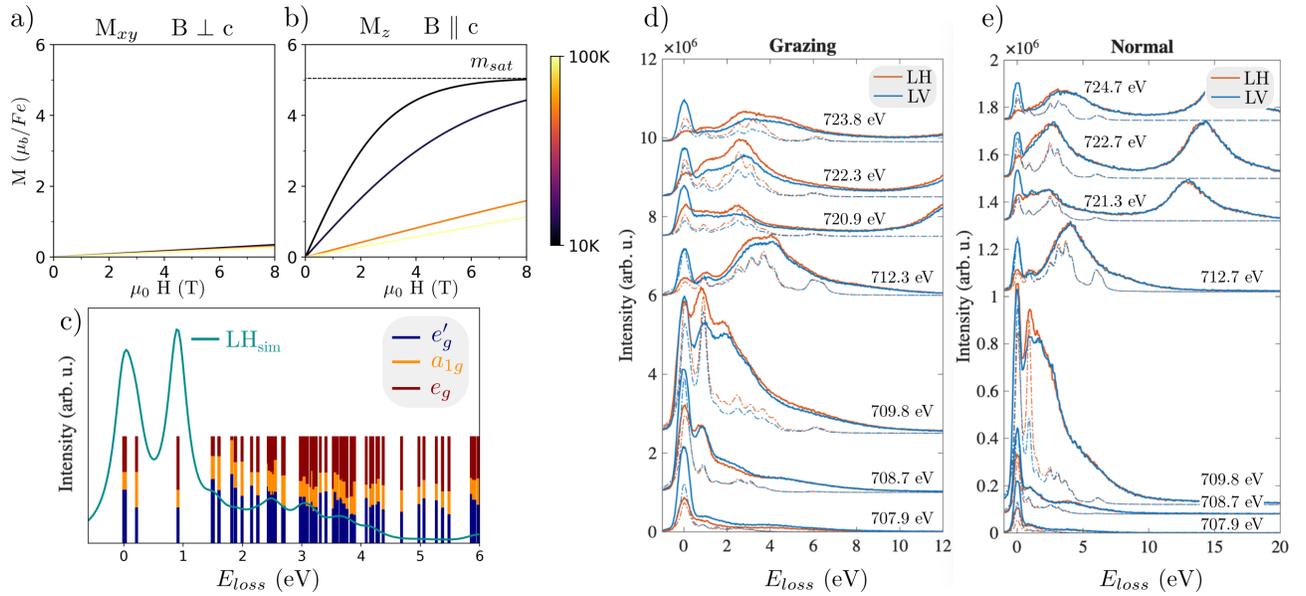

FIG. S1. Magnetization curves with field perpendicular in a) and parallel in b) to the crystallographic $c$ axis showing the role of anisotropy. c) Computed energy level projected on the character of the orbitals superimposed with a RIXS spectrum. d) and e) are additional RIXS spectra and simulation thereof in grazing and normal geometry.

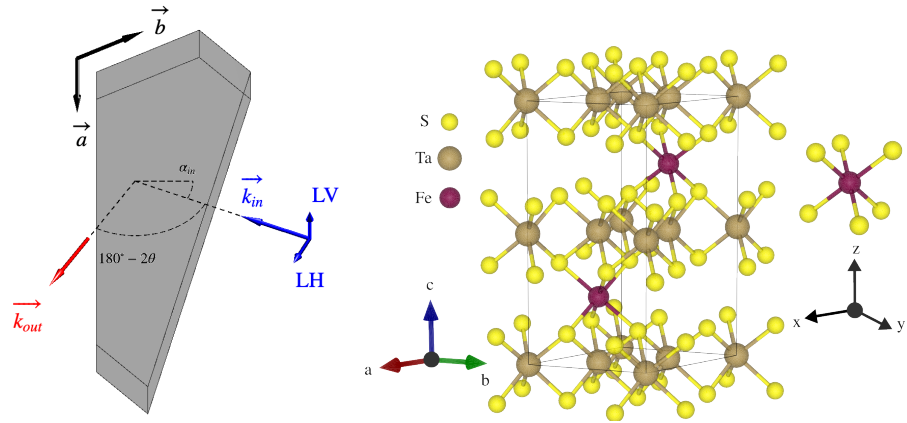

FIG. S2. a) Scheme of the experimental geometry presenting the orientation of the incident and outgoing radiation with respect to the crystal and its crystallographic axes. b) Crystallographic unit cell of $Fe_{1/3}TaS_2$ with Fe, Ta and S atoms depicted in violet, khaki and yellow, respectively. The local xyz reference frame used for the model calculation is also included, having its x and z axes collinear with the crystallographic a and c directions.

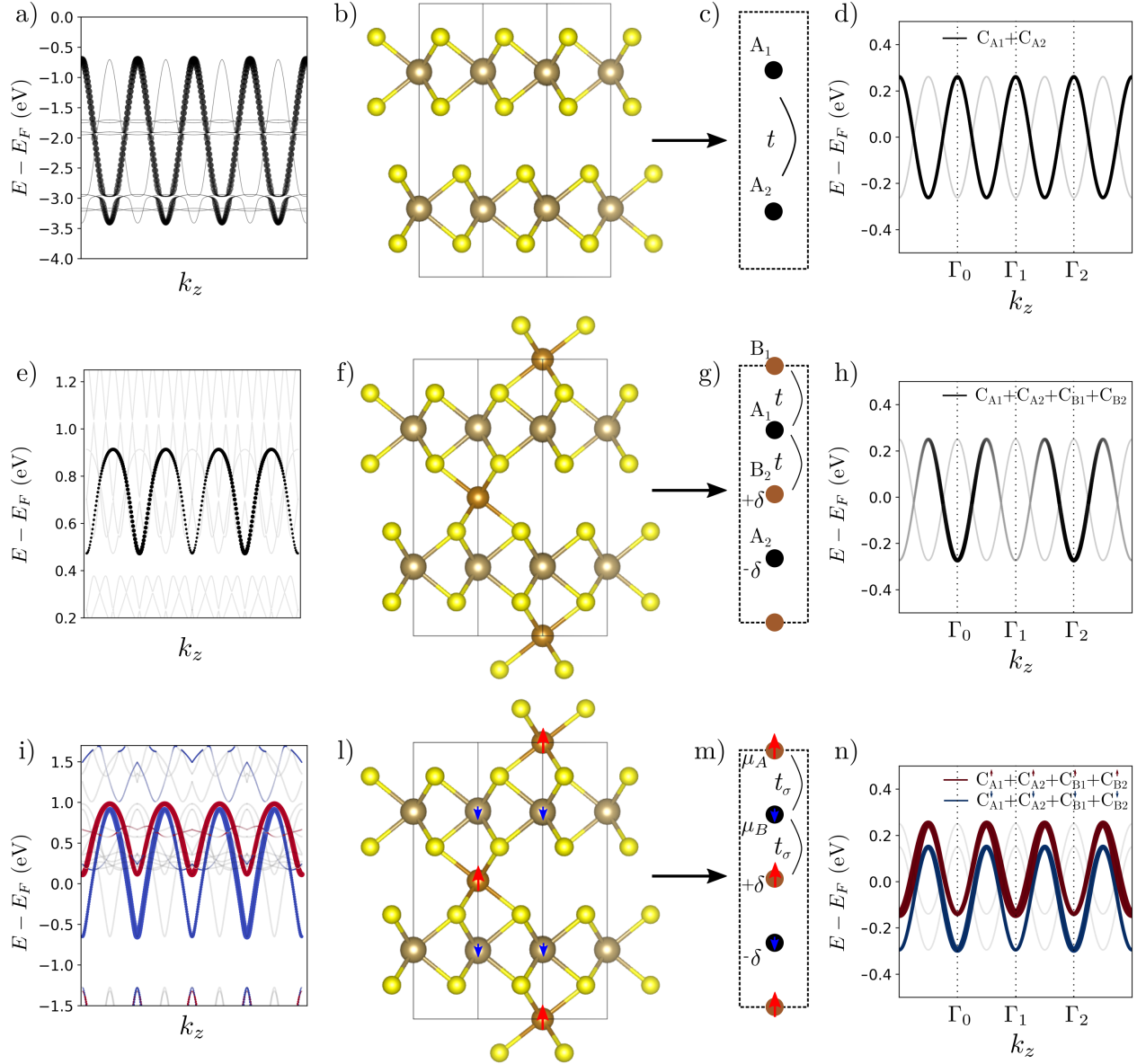

FIG. S3. a)-e)-i) First principle out-of-plane dispersion of pristine 2H-TaS$_2$, non-magnetic Fe$_{1/3}$TaS$_2$ and ferromagnetic Fe$_{1/3}$TaS$_2$, respectively. b)-f)-l) Magnetic and structural unit cell of the corresponding phases. c)-g)-m) Sketch of one dimensional models capturing the $k_z$ dispersion. d)-h)-n) Band structure of the one dimensional model.

## C. ARPES and SX-ARPES

ARPES experiments were performed at the Bloch beamline of the MAX IV synchrotron (Sweden) using DA30-L ScientaOmicron analyzer with a horizontal slit. The photon energies used were in the 75 and 130 eV range with linear horizontal polarized light (LH polarization) and the sample temperature was T = 15 K. The $k_z$ dispersion is calculated using an inner potential of 10 eV. The samples, commercial crystals from HQ graphene, were exfoliated in the UHV chamber using a Kapton tape. Preliminary ARPES data were also collected at VUV-beamline of Elettra synchrotron (Trieste, Italy) using a Scienta R4000-WAL analyzers with a horizontal slit. These latter data are not reported in the present study for the sake of conciseness. SX-ARPES experiments were performed at the P04 beamline of the PETRAIII synchrotron at DESY (Germany) delivering circular polarized light. The endstation, migrated from the ADRESS beamline of the Swiss Light Source and based on a grazing-incidence experimental geometry, used a

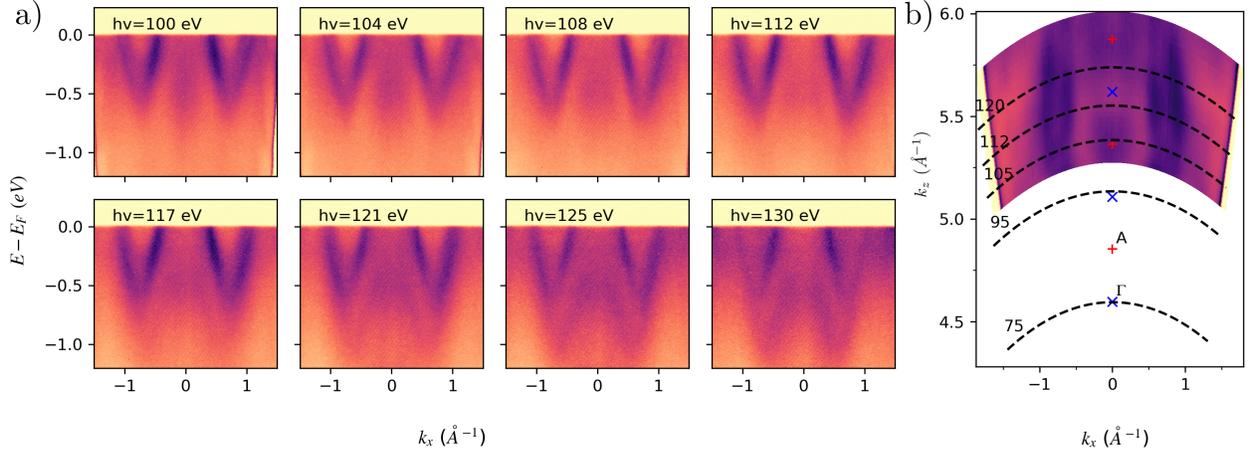

FIG. S4. a) In-plane dispersion along the $\Gamma K$ direction at different photon energy. b) Fermi surface in the $\Gamma K A$ plane reporting different lines indicating constant photon energy curves.

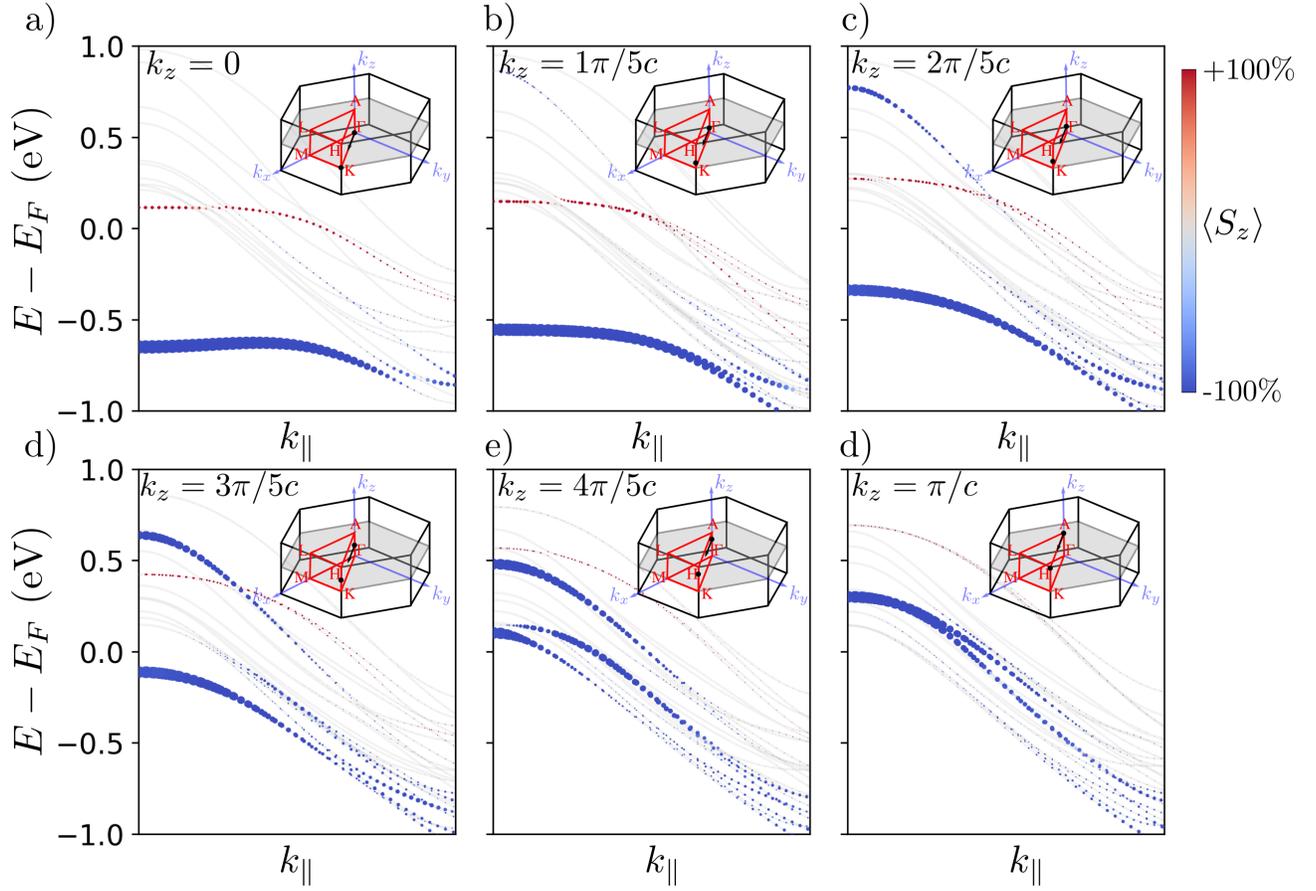

FIG. S5. In-plane dispersion at different $k_z$. We also report the Brillouin Zone and the path chosen for the band structure. The size of the points is proportional to the Fe-$d$ character, while the color is the spin-polarization.

PHOIBOS225 analyzer with a vertically oriented slit. The measurements used photon energies in the range 350-700 eV. The combined energy resolution varied between 50 and 100 eV, respectively. The sample temperature was T = 15 K. Low energy electron diffraction (LEED) showed the $\sqrt{3} \times \sqrt{3} R30°$ pattern with respect to the pristine 2H-TaS$_2$ typical of the $x = 1/3$ concentration (see Fig. 1c).

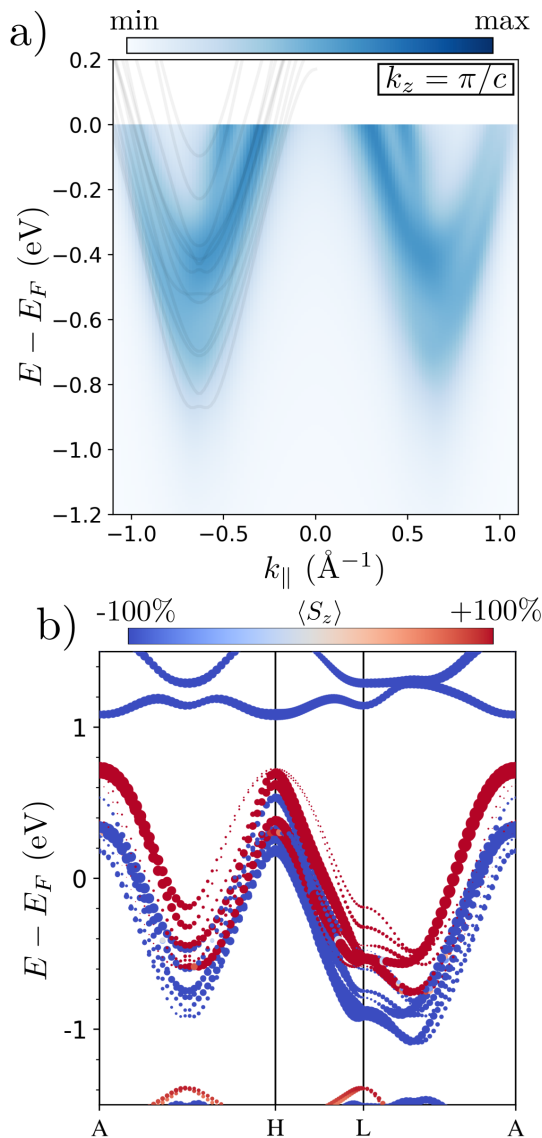

FIG. S6. a) Simulation of the ARPES intensity. On the left, we superimposed the band structure calculated using the PBE+$U$+$SOC$ functional. b) Unfolded and spin-polarized band structure. The size of the point are proportional to the weight of the unfolding, while the colormap is for the spin polarization $\langle S_z \rangle$ in the AHL plane.

## III. MODELS FOR $k_z$ DISPERSION

In order to describe the $k_z$ dispersion of 2H-TaS$_2$ and intercalated Fe$_{1/3}$TaS$_2$ we report in Fig. S3 first principle calculation and one dimensional models capturing its main features. In Fig.S3 panel a we report first-principle calculation where the size of the points is proportional to the square modulus of $M \sim \sum_{S_j} C_{S_j}^{n\mathbf{k}}$ as in Ref. [11], where $S_j$ is S-$p_z$ orbitals. This band dispersion can be described by one dimensional monoatomic chain as shown in Fig. S3 panel d where the weight is proportional to the sum of A1 and A2 sites. First-principle calculation of non-magnetic Fe$_{1/3}$TaS$_2$ (Fig. S3 panel e) show that the weight, proportional to the square modulus of $M \sim \sum_{Ta_j} C_{Ta_j}^{n\mathbf{k}} + \sum_{Fe_j} C_{Fe_j}^{n\mathbf{k}}$, shows a different modulation along $k_z$ with respect to the pristine case (where $Ta_j$ and $Fe_j$ represent Ta-$d^{z^2}$ and Fe-$d^{z^2}$ orbitals). This feature is well captured by a monoatomic chain with two atoms with different masses (see sketch in Fig. S3 panel g). The band structure weighted with $|C_{A1} + C_{A2} + C_{B1} + C_{B2}|^2$ well describe the $k_z$ modulation, demonstrating that it arises from a pure geometric contribution to the Bloch phase. Finally, to describe the dispersion of the realistic ferromagnetic Fe$_{1/3}$TaS$_2$ we constructed a spinful tight-binding model with different Zeeman $\mu_i$ ($i = A, B$), on-sites ($\pm\delta$) (see Fig. S3 panel i-l-m-n) and spin dependent hoppings $t_\sigma = \sigma t$ with $\sigma = \pm 1$.

## IV. COMPLEMENTARY DATA AND SIMULATIONS

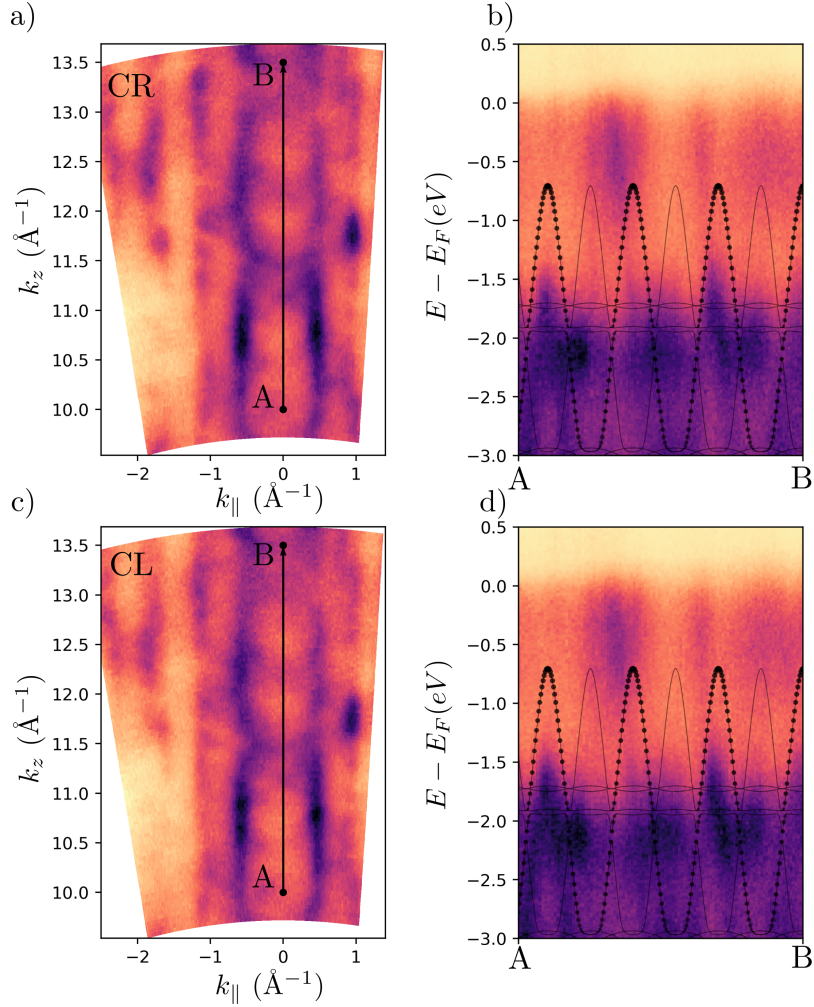

FIG. S7. a) Fermi surface in the $\Gamma AK$ plane using CR and b) corresponding band structure at $(k_x, k_y)=0$ in which we superimpose band structure of the pristine 2H-TaS$_2$. The same for CL light in c) and d).

In Fig. S4 panel a we report the in-plane dispersion along the $\Gamma K$ direction at different photon energy and in panel b the Fermi-surface in the $\Gamma KA$ with photon energy ranging from 100 to 130 eV.

To understand the in-plane dispersion of the Fe-$d$-Ta-$d$ bands as a function of the out-of-plane momentum, we plot the band structure along the parallel direction as a function of the out-of-plane momentum (see Fig. S5). To clearly distinguish the Fe hybridized bands, the size of the points is proportional to the Fe-$d$ character of the band. As reported in Fig. 4, while the dispersion along $k_z$ is nearly linear around the Fermi level, along $k_{||}$ is parabolic as evident by the cuts at different $k_z$'s in Fig. S5. This strongly anisotropic behavior is typical of semi-Dirac materials where the dispersion along different $k$ directions is qualitative different [26–29]. In particular, in our 3D case, the energy dispersion near the Fermi level can be written as $E(k_x, k_y, k_z) = k_x^2/2m_1 + k_y^2/2m_2 + v_F k_z$ with $m_1$ and $m_2$ two different effective masses.

In Fig.S6, panel a, we present the simulated ARPES intensity at $k_z = \pi/c$ together with the corresponding unfolded band structure along the high-symmetry lines in the AHL plane. When comparing this unfolded dispersion with that in the $\Gamma KM$ plane, we find that the splittings associated with the 2H stacking symmetry disappear: the bands that were previously split now become degenerate in the AHL plane. Importantly, the Zeeman splitting of the Ta $d$-derived bands induced by ferromagnetic Fe intercalation remains intact.

Finally, in Fig. S7, we show a comparison of the $k_z$ dispersion acquired with circular right (CR) and circular left light (CL) , showing no substantial differences. We superimposed to the SX-ARPES spectrum the out-of-plane dispersion of pristine 2H-TaS$_2$ to directly demonstrate how the out-of-plane dispersion is strongly modified by the Fe intercalant atom.